\documentclass[%
a4paper,
reprint,
superscriptaddress,
nofootinbib,
longbibliography,
amsmath,amssymb,
aps, prd,
]{revtex4-2c/revtex4-2}

\usepackage{graphicx}% Include figure files
\usepackage[caption=false]{subfig}
\usepackage{placeins}
\renewcommand{\thefootnote}{\fnsymbol{footnote}}% use symbols as footnotes
\usepackage{dcolumn}% Align table columns on decimal point
\usepackage{multirow}% cell spans multiple rows
\usepackage{bm}% bold math
\usepackage[shortcuts]{extdash}% for protected dash \=/
\usepackage[dvipsnames,svgnames,x11names,hyperref]{xcolor}
\usepackage[colorlinks]{hyperref}% add hypertext capabilities
\usepackage{mathtools}% for defining \Pr
\usepackage{physics} % write derivatives as \dv or \dvp
\usepackage{siunitx} % use this package module for SI units
\usepackage[capitalise,nameinlink]{cleveref} %Referencing without need to explicitly state fig /table

\newcommand{\e}{\,\mathrm{e}}
\defcitealias{Planck2018Parameters}{P18}

\DeclareSIUnit\clight{\mathit{c}}
\DeclareSIUnit\astronomicalunit{au}
\DeclareSIUnit\littleh{\mathit{h}}
\DeclareSIUnit\nat{nat}
\DeclareSIUnit\nats{nats}
\DeclareSIUnit\parsec{pc}
\DeclareSIUnit\efolds{e\text{-}folds}
\DeclareSIUnit\planckmass{m_\mathrm{p}}
\DeclareSIUnit\plancktime{t_\mathrm{p}}
\DeclareSIUnit\plancklength{\ell_\mathrm{p}}

\begin{document}

\preprint{APS/123-QED}

\title{Bayesian evidence for the tensor-to-scalar ratio~\texorpdfstring{$r$}{ } and neutrino masses~\texorpdfstring{$m_\nu$}{ }:\texorpdfstring{\\}{ } Effects of uniform vs logarithmic priors}% Force line breaks with \\

\author{L.~T.~Hergt}%
    \email{lh561@mrao.cam.ac.uk}%
\author{W.~J.~Handley}%
    \email{wh260@mrao.cam.ac.uk}%
    \affiliation{Astrophysics Group, Cavendish Laboratory, J.~J.~Thomson Avenue, Cambridge, CB3~0HE, UK}%
    \affiliation{Kavli Institute for Cosmology, Madingley Road, Cambridge, CB3~0HA, UK}%
\author{M.~P.~Hobson}%
    \email{mph@mrao.cam.ac.uk}%
    \affiliation{Astrophysics Group, Cavendish Laboratory, J.~J.~Thomson Avenue, Cambridge, CB3~0HE, UK}%
\author{A.~N.~Lasenby}%
    \email{a.n.lasenby@mrao.cam.ac.uk}%
    \affiliation{Astrophysics Group, Cavendish Laboratory, J.~J.~Thomson Avenue, Cambridge, CB3~0HE, UK}%
    \affiliation{Kavli Institute for Cosmology, Madingley Road, Cambridge, CB3~0HA, UK}%
    
\date{\today}% It is always \today, today, but any date may be explicitly specified

\begin{abstract}
We review the effect that the choice of a uniform or logarithmic prior has on the Bayesian evidence and hence on Bayesian model comparisons when data provide only a one-sided bound on a parameter. 
We investigate two particular examples: the tensor-to-scalar ratio\texorpdfstring{~$r$}{} of primordial perturbations and the mass of individual neutrinos\texorpdfstring{~$m_\nu$}{}, using the cosmic microwave background temperature and polarisation data from Planck~2018 and the NuFIT~5.0 data from neutrino oscillation experiments. 
We argue that the Kullback--Leibler divergence, also called the relative entropy, mathematically quantifies the Occam penalty.
% and demonstrate in a mock example how the Bayesian log-evidence is traded in for relative entropy when switching between uniform and logarithmic priors.
We further show how the Bayesian evidence stays invariant upon changing the lower prior bound of an upper constrained parameter.
% We confirm this in our analysis of the tensor-to-scalar ratio and of the lightest neutrino mass. 
% Using a logarithmic prior instead of a uniform prior frees the $\Lambda$CDM extensions from their additional Occam penalty bringing the Bayesian evidence up to par with the $\Lambda$CDM model without extension.
While a uniform prior on the tensor-to-scalar ratio disfavours the $r$-extension compared to the base $\Lambda$CDM model with odds of about $1:20$, switching to a logarithmic prior renders both models essentially equally likely.
$\Lambda$CDM with a single massive neutrino is favoured over an extension with variable neutrino masses with odds of $20:1$ in case of a uniform prior on the lightest neutrino mass, which decreases to roughly $2:1$ for a logarithmic prior.
% As long as the lower bound of the logarithmic priors is chosen sufficiently small where the likelihood becomes invariant, then the Bayesian evidence becomes invariant as well.
% Using neutrino oscillation data as prior information on the neutrino mass splitting, the choice of prior on the lightest neutrino mass does not affect the comparison of neutrino hierarchies. 
For both prior options we get only a very slight preference for the normal over the inverted neutrino hierarchy with Bayesian odds of about $3:2$ at most.
% Quoting one-sided posterior bounds in terms of percentiles works better for uniform priors. For logarithmic priors a percentile bound would depend on the unconstrained prior bound. Instead we recommend determining the step-position where the posterior has dropped to a fraction of its maximum value.
\end{abstract}

%Use showkeys class option if keyword display desired

\maketitle

\section{Introduction} 

The ``principle of insufficient reason'' (Bernoulli~\cite{Bernoulli1713}) or ``principal of indifference'' (renamed by Keynes~\cite{Keynes1921}) states that in the event of multiple, mutually exclusive, possible outcomes and in the absence of any relevant evidence, we should assign the same probability to all outcomes~\cite{Sivia2006Book}. In a Bayesian analysis, this is generalised to continuous parameters in the form of uninformative priors. Complete prior ignorance about a location parameter is represented by assigning a uniform distribution to the prior. Ignorance about a scale parameter on the other hand is represented by assigning a logarithmic prior, i.e.\ a uniform distribution on the logarithm of the parameter~\cite{Sivia2006Book}. 
However, it is not always clear whether a parameter should be treated as a location or scale parameter. This is quite commonly discussed when faced with a strictly positive parameter such as a mass or an amplitude that is very small, yet still unconstrained. 
In general, the decision whether to use a uniform or logarithmic prior has effects on credibility bounds and on the Bayesian evidence, i.e.\ on both levels of Bayesian inference: parameter estimation and model comparison.
Under the reasoning that you can set the lower bound to zero and thus incorporate all possible small values, the uniform prior is often preferred, whereas the logarithmic prior is criticised for a lack of an unambiguous lower bound, and because the ultimate choice of the lower bound might affect a \SI{95}{\percent} credibility bound and the Bayesian evidence.
 
In this paper we show that the very last statement is typically not true and that the choice of a lower bound for such a logarithmic prior is less problematic than commonly assumed. To that end we will look at two cosmological examples in particular: the tensor-to-scalar ratio~$r$ of primordial perturbations as well as the neutrino masses~$m_\nu$, where both uniform and logarithmic priors have been applied historically (for the tensor-to-scalar ratio, see e.g.~\cite{Lau2014, Barenboim2015, Creminelli2015, Planck2018Parameters, BicepKeck2018BKX, Hirano2019}; and for the neutrino masses see e.g.~\cite{Simpson2017, Schwetz2017, Caldwell2017, Capozzi2017, Planck2018Parameters, Loureiro2019, Capozzi2020, Archidiacono2020, Choudhury2020, Stoecker2020}). 
 
The best constraints on the tensor-to-scalar ratio $r_{0.05}\!\lesssim\!0.06$ come from joint analyses of cosmic microwave background~(CMB) data, CMB lensing, and baryon acoustic oscillations~(BAO)~\cite{Planck2018Parameters, BicepKeck2018BKX}, where a uniform prior on~$r$ was adopted.
A common goal of upcoming CMB experiments such as the Simons Observatory~\cite{SimonsObservatoryScience}, the LiteBIRD satellite~\cite{LiteBird2020WhitePaper} and the next-generation ``Stage\=/4'' ground-based CMB experiment (CMB\=/S4)~\cite{CMBS4Science} is to push to a tensor-to-scalar ratio of $r \sim 10^{-3}$. In pushing to such small values of~$r$, the question of whether to adopt a uniform or logarithmic prior in one's analysis becomes more pertinent.

Since neutrino oscillation experiments measure non-zero mass differences, we can conclude that two or more neutrinos must have mass. However, the absolute scale of the individual neutrino masses~$m_i$ cannot be measured by the oscillation experiments, but only the mass-squared splittings $\Delta m_{ij}^2 = m_i^2 - m_j^2$.
The strongest bound on the absolute neutrino mass scales is currently provided again by combined CMB and BAO data, limiting the sum of the neutrino masses to $\sum m_\nu \lesssim \SI{0.12}{\eV}$ at \SI{95}{\percent} confidence~\cite{Choudhury2020} (see also \cite{Capozzi2020,Stoecker2020} for other recent analyses).

When investigating the three discrete neutrino mass eigenstates, the question of uniform vs logarithmic priors arises again. Note, however, that given the known mass splittings from oscillation experiments, the three neutrino mass scales are linked. If one mass scale is known, then the others can be inferred from the mass squared splittings. Hence, only one mass scale is truly unknown and assuming scale invariant (i.e.\ logarithmic) priors on all three neutrino masses simultaneously would unduly favour smaller neutrino masses and thus a normal neutrino hierarchy~(NH) with $m_1 < m_2 \ll m_3$ compared to an inverted neutrino hierarchy~(IH) with $m_3 \ll m_1 < m_2$ (for more on this see also discussions in~\cite{Simpson2017,Schwetz2017}).

This paper is structured as follows: In \cref{sec:methods} we will start by giving a brief description of our Bayesian analysis framework, including the data and base cosmological model used, as well as the means of computing the Bayesian evidence. In \cref{sec:r} we apply this to the tensor-to-scalar ratio~$r$ and compare to a theoretical mock example. In \cref{sec:mnu} we perform the equivalent analysis for the neutrino masses and contrast the results for the two neutrino hierarchies. We conclude in \cref{sec:discussion}.

\section{Methods}
\label{sec:methods}

\subsection{Bayesian inference}
\label{sec:bayes}

There are two levels to Bayesian inference: parameter estimation and model comparison (see e.g.~\cite{Sivia2006Book,MacKay2003ch28}).
Both these levels are based on Bayes' theorem which relates inference inputs (likelihood and prior) to yielded outputs (posterior and evidence):
\addtolength{\jot}{6pt}
\begin{alignat}{3}
    \Pr(\theta|D,M)       &\times \Pr(D|M)         & &=\,\, & \Pr(D|\theta,M)       &\times \Pr(\theta|M) , \nonumber \\
    \mathrm{Posterior}    &\times \mathrm{Evidence}& &=\,\, & \mathrm{Likelihood}   &\times \mathrm{Prior} , \nonumber \\
    \label{eq:bayes}
    \mathcal{P}_M(\theta) &\times \mathcal{Z}_M    & &=\,\, & \mathcal{L}_M(\theta) &\times \pi_M(\theta) . 
\end{alignat}
\setlength{\jot}{3pt}
The posterior~$\mathcal{P}$ is the main quantity of interest in a parameter estimation, representing our state of knowledge about the parameters~$\theta$ in a given model~$M$, inferred from our prior information~$\pi$ and the likelihood~$\mathcal{L}$ of the parameters under the data~$D$. The evidence~$\mathcal{Z}$ is pivotal for model comparisons.

Were we interested only in parameter estimation, then it would be sufficient to care only about the proportionality of the posterior to the product of likelihood and prior and the Bayesian evidence could be neglected as a mere normalisation factor. However, for the comparison of say two models~$A$ and~$B$ the evidence becomes important with the posterior odds ratio of the two models given by: 
\begin{align}
    \label{eq:posterior_ratio}
    \frac{\Pr(B|D)}{\Pr(A|D)} = \frac{\Pr(B)}{\Pr(A)} \times \frac{\mathcal{Z}_B}{\mathcal{Z}_A} .
\end{align}
Typically models are assigned the same prior preference such that the first term on the right-hand side becomes unity, leaving simply the evidence ratio $\mathcal{Z}_B/\mathcal{Z}_A$, which can be interpreted as betting odds for the two models. We typically quote this in terms of the log-difference of evidences between two models $\Delta\ln\mathcal{Z} = \ln(\mathcal{Z}_B/\mathcal{Z}_A)$.

The evidence is the marginal likelihood
\begin{equation}
    \mathcal{Z}_M = \int \mathcal{L}_M(\theta) \, \pi_M(\theta) \,\dd\theta = \big\langle \mathcal{L}_M \big\rangle_\pi ,
\end{equation}
and can be numerically approximated with Laplace's method~\cite{MacKay2003ch27}, estimated from a posterior distribution attained e.g.\ from a Monte Carlo Markov Chain~(MCMC) via the Savage--Dickey density ratio~(SDDR)~\cite{Dickey1971,Trotta2007,Trotta2007a,Verde2013} or via a nearest-neighbour approach~\cite{Heavens2017,Heavens2017a} or computed more directly with nested sampling, which can additionally estimate the corresponding numerical uncertainty~\cite{Skilling2006,Sivia2006ch9,MultiNest1,MultiNest2,MultiNest3,PolyChord1,PolyChord2}. 

If the posterior distribution and the evidence have both been determined, then as a byproduct one can also compute the Kullback--Leibler~(KL) divergence, also called the relative entropy:
\begin{equation}
    \label{eq:DKL}
    \mathcal{D}_{\mathrm{KL},M} = \int \mathcal{P}_M(\theta) \ln\left( \frac{\mathcal{P}_M(\theta)}{\pi_M(\theta)} \right) \dd\theta 
    = \left\langle \ln\frac{\mathcal{P}_M}{\pi_M} \right\rangle_\mathcal{P} ,
\end{equation}
which quantifies the overall compression from prior to posterior distribution. 

\subsection{Kullback--Leibler divergence and Occam's razor}
\label{sec:DKL}

It should be noted that the Bayesian evidence naturally incorporates the so-called Occam's razor that penalises models for unnecessary complexity. It can be formulated as the principle to ``Accept the simplest explanation that fits the data''~\cite{MacKay2003ch28}.
This can be neatly demonstrated using a Gaussian likelihood with mean~$\mu$ and variance~$\sigma^2$ having a single parameter $x\in[x_\mathrm{min}, x_\mathrm{max}]$ with a uniform prior (see e.g.~\cite{Sivia2006ch4,MacKay2003ch28}). The Bayesian evidence decomposes into two terms:
\begin{equation}
    \label{eq:MrAandMrB}
    \mathcal{Z} = \mathcal{L}(\mu) \times \frac{\sigma \sqrt{2\pi}}{x_\mathrm{max}-x_\mathrm{min}} 
\end{equation}
The first term on the right-hand side is the maximum likelihood point. With additional parameters, this term would only increase and therefore can only favour the given model. The second term incorporates the ratio of posterior to prior uncertainty. Since the posterior uncertainty~$\sigma$ is generally smaller than the prior uncertainty $(x_\mathrm{max}-x_\mathrm{min})$, this term penalises the given model for each of its parameters and thus embodies its Occam penalty. Note that the posterior and prior uncertainties appear \emph{inversely} in the normalisation factor of the actual distributions. 

More generally, the KL-divergence can actually be used as an estimator of the Occam penalty, which becomes clearer when rewriting the log-evidence according to\footnote{Note that proving \cref{eq:evidence_fit_occam} becomes surprisingly straight-forward when going from right to left and making use of Bayes' theorem~\labelcref{eq:bayes}.}:
\addtolength{\jot}{6pt}
\begin{alignat}{3}
\label{eq:evidence_fit_occam}
    \ln( \int\!\! \mathcal{L}_M \pi_M \dd\theta\!) \! &= & 
    \int\! \mathcal{P}_M \ln\mathcal{L}_M \dd\theta &-\! 
    \int\! \mathcal{P}_M \ln(\frac{\mathcal{P}_M}{\pi_M}) \dd\theta , \nonumber \\
    \text{(log-)evidence} &= & \text{parameter fit}\,   &- \,\text{Occam penalty} ,\nonumber \\
    \ln\mathcal{Z}_M  &= & \langle \ln\mathcal{L}_M \rangle_\mathcal{P} &- \mathcal{D}_{\mathrm{KL},M},
\end{alignat}
where we have dropped the dependence on~$\theta$ to save space. Analogous to the example from \cref{eq:MrAandMrB}, the first term on the right-hand side encapsulates the fit of the model, while the KL-divergence is the average log-ratio of posterior to prior distribution (see also the last equality in \cref{eq:DKL}), thus identifying it as the Occam penalty.
% {\color{RoyalBlue} This has been intuitively applied e.g.\ in the third figure in~\cite{Handley2019c}, but as far as the authors are aware \cref{eq:evidence_fit_occam} is the first time this has been stated analytically in this form.
% }
This equation appears in passing in the appendix of~\cite{Heymans2020} in the calculation of tension metrics and it has been intuitively applied e.g.\ in the third figure in~\cite{Handley2019c}, but as far as the authors are aware \cref{eq:evidence_fit_occam} is the first time this analytical form is explicitly connected to the trade-off between parameter fit and model complexity.

While known to experts, a widely unappreciated fact is that the evidence stays unaffected by an \emph{un}constrained parameter, i.e.\ when the data provide no information for that parameter~\cite{Handley2019b}. 
In terms of \cref{eq:evidence_fit_occam} this is reflected in an invariant likelihood $\langle\ln\mathcal{L}\rangle_\mathcal{P}$ and a zero KL-divergence $\mathcal{D}_\mathrm{KL}(\pi\!=\!\mathcal{P})\!=\!0$. Using the alternative labelling we can rephrase this: Adding an unconstrained parameter does not affect the fit, but also does not incur an additional Occam penalty and hence also leaves the evidence unaffected.

A popular measure for an effective number of constrained parameters is the Bayesian model complexity~\cite{Spiegelhalter2002}. However, this quantity relies on the use of a point estimator such as the posterior mean or mode, which is why we prefer using the Bayesian model dimensionality~$d$ in the following sections (see~\cite{Handley2019a} for a more detailed discussion on Bayesian complexities/dimensionalities). The Bayesian model dimensionality can be computed straightforwardly from the posterior distribution as the posterior \emph{variance} of the log-likelihood:
\begin{align}
    \frac{d_M}{2} 
    &= \int \mathcal{P}_M(\theta) \left( \ln\frac{\mathcal{P}_M(\theta)}{\pi_M(\theta)} - \mathcal{D}_{\mathrm{KL},M} \right)^2 \dd{\theta} \\
    &= \big\langle \left( \ln\mathcal{L}_M \right)^2 \big\rangle_\mathcal{P} - \big\langle \ln\mathcal{L}_M \big\rangle_\mathcal{P}^2 . 
    %= \mathrm{Var}_\mathcal{P}(\ln\mathcal{L}_M) .
\end{align}
Note the connection to \cref{eq:evidence_fit_occam}, where we used the posterior \emph{average} of the log-likelihood.
As such, these two quantities provide an interesting additional perspective to that of the $(\ln\mathcal{Z}, \mathcal{D}_\mathrm{KL})$ pair. The posterior average of the log-likelihood informs us about the parameter fit and the posterior variance of the log-likelihood measures the models' complexity in the form of the number of constrained parameters.

\subsection{Cosmological models}

In the following sections we perform Bayesian model comparisons on one-parameter extensions to the $\Lambda$CDM model (universe dominated today by a cosmological constant~$\Lambda$ and by cold dark matter), which we parametrise with the standard~6 cosmological parameters listed in \cref{tab:prior} with their corresponding prior ranges.

In \cref{sec:r} we extend the $\Lambda$CDM model by the tensor-to-scalar ratio~$r$ of primordial perturbations, which is set to $r=0$ in $\Lambda$CDM. In \cref{sec:mnu} we extend the base model by allowing for three distinct neutrino masses. In the $\Lambda$CDM model these are typically fixed to two massless neutrinos and a single massive neutrino with $m_\nu=\SI{0.06}{\eV}$.

\newcolumntype{+}{D{,}{\,\pm\,}{-1}}
\newcolumntype{,}{D{,}{,}{-1}}
\begin{table}[tb]
\renewcommand{\arraystretch}{1.5}
    \caption{\label{tab:prior} Cosmological parameters of the base $\Lambda$CDM cosmology the way they are sampled in our Bayesian analysis. The second column shows their corresponding prior ranges. The third column lists their mean and \SI{68}{\percent} limits from our base $\Lambda$CDM nested sampling run with TT,TE,EE+lowE data from Planck~2018 and is in almost perfect agreement with table~2 in~\cite{Planck2018Parameters}.}
\begin{ruledtabular}
    
\begin{tabular}{ l c + }
    Parameter & Prior range & \multicolumn{1}{c}{\SI{68}{\percent} limits} \\ \hline
    $\omega_\mathrm{b}=h^2\Omega_\mathrm{b}$ & [0.019, 0.025] & 0.02236 , 0.00015        \\
    $\omega_\mathrm{c}=h^2\Omega_\mathrm{c}$ & [0.025, 0.471] & 0.1199  , 0.0014         \\
    $100\,\theta_\mathrm{s}$                 & [1.03, 1.05]   & 1.04191 , 0.00029        \\
    $\tau_\mathrm{reio}$                     & [0.01, 0.40]   & 
    \multicolumn{1}{c}{$\hphantom{0\,} 0.0540_{\,-\,0.0084}^{\,+\,0.0073}$} \\
    $\ln(10^{10} A_\mathrm{s})$              & [2.5, 3.7]     & 
    \multicolumn{1}{c}{$\hphantom{0\,} 3.043_{\,-\,0.016}^{\,+\,0.015}$}    \\
    $n_\mathrm{s}$                           & [0.885, 1.040] & 0.9641  , 0.0042         \\
\end{tabular}

\end{ruledtabular}
\end{table}

\subsection{Data}

We use the 2018 temperature and polarisation data from the Planck satellite~\cite{Planck2018CMB}, which we abbreviate as ``TT,TE,EE+lowE''.
Note that this is the same abbreviation as in the corresponding Planck publication itself. The specific use of ``lowE'' but lack of ``lowT'' might lead to the conclusion that only E-mode and no temperature data were used at low multipoles. However, this is \emph{not} the case. Both high-$\ell$ \emph{and} low-$\ell$ temperature auto-correlation data are implied in that abbreviation.

In \cref{sec:mnu} we additionally use the NuFIT~5.0~(2020) data from neutrino oscillation experiments~\cite{Esteban2019,Esteban2020,NuFIT} to set Gaussian priors on the mass squared splittings~$\delta m^2$ and~$\Delta_m^2$.

\begin{figure*}[tb]
    \includegraphics[scale=0.95]{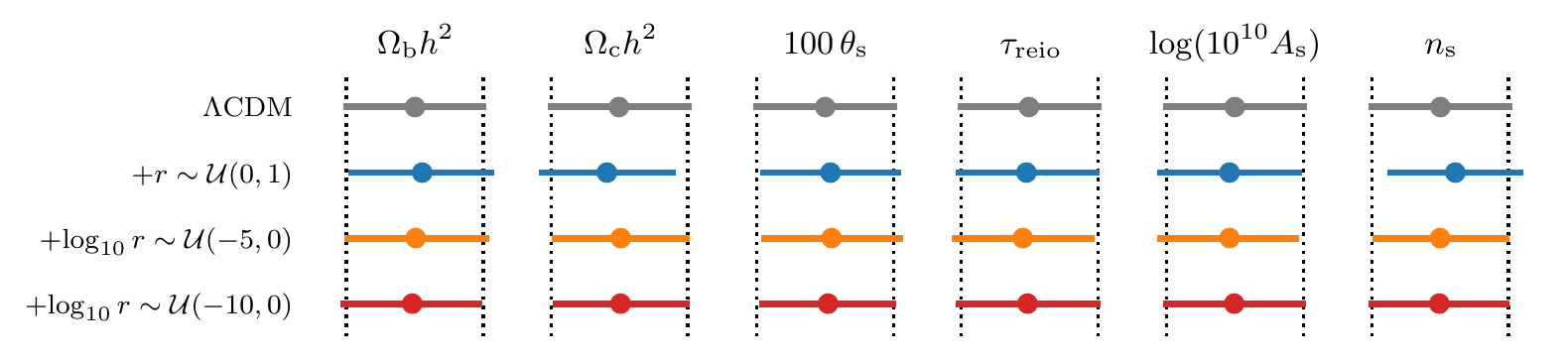}
    \caption{\label{fig:r_param_stability} Stability of the cosmological parameters for the tensor-to-scalar ratio extension of the base $\Lambda$CDM cosmology with different priors on~$r$: uniform in blue, logarithmic with lower bound~$-5$ in orange and logarithmic with lower bound~$-10$ in red. For each parameter we show the mean and the extent from quantile $0.16$ to $0.84$, i.e.\ the inner \SI{68}{\percent} limits.}
\end{figure*}

\subsection{Statistical and cosmological software}

We explore the posterior distributions of cosmological and nuisance parameters using \texttt{Cobaya}~\cite{Cobaya1}, which provides both the MCMC sampler developed for \texttt{CosmoMC}~\cite{CosmoMC1,CosmoMC2} with a ``fast dragging'' procedure described in~\cite{CosmoMC3} and also the nested sampling code \texttt{PolyChord}~\cite{PolyChord1,PolyChord2}, tailored for high-dimensional parameter spaces, which can simultaneously determine the Bayesian evidence alongside its numerical uncertainty.
Both samplers are interfaced with the cosmological Boltzmann code \texttt{CLASS}~\cite{Class1,Class2,Class4}, which computes the theoretical CMB power spectra for temperature and polarisation modes. 

We use~\texttt{GetDist}~\cite{GetDist} to generate the data tables of marginalised parameter values.
The post-processing of the nested sampling output for the computation of Bayesian evidence, KL-divergence and Bayesian model dimensionality, as well as the plotting functionality for posterior contours is performed using the python module \texttt{anesthetic}~\cite{Anesthetic}. 

All inference products required to compute the results presented in this paper are available for download from \texttt{Zenodo}~\cite{ZenodoUniLog}.

\section{Tensor-to-scalar ratio}
\label{sec:r}

The tensor-to-scalar ratio~$r$ quantifies what fraction of primordial perturbations is in the form of gravitational waves, produced e.g.\ during cosmic inflation and potentially detectable in their contribution to CMB B\=/modes. 

So far, the major experiments probing the contribution of tensor modes to the CMB power spectrum have adopted a uniform prior on~$r$~\cite{Planck2018Parameters,BicepKeck2018BKX}. However, the common target of $r \sim 10^{-3}$ for many upcoming CMB experiments such as the Simons Observatory, the LiteBIRD satellite or CMB\=/S4, warrants the question as to whether a scale invariant prior might be better to handle such low values. This question frequently brings up arguments of the ambiguity of the lower bound to a logarithmic prior and its potential effect on the Bayesian evidence.

\subsection{Tensor-to-scalar ratio: Posteriors}
\label{sec:r_posterior}

\Cref{fig:r_param_stability} gives an overview of the stability of the cosmological base parameters across different priors for~$r$ and compares them to the $\Lambda$CDM base model by showing their mean and \SI{68}{\percent} ranges.
In addition to the $\Lambda$CDM base run, we have taken nested sampling runs with both a uniform prior on the tensor-to-scalar ratio $r\!\sim\!\mathcal{U}(0,1)$ and with two logarithmic priors with different lower bounds, $\log_{10}r\!\sim\!\mathcal{U}(-5,0)$ and $\log_{10}r\!\sim\!\mathcal{U}(-10,0)$.
The near perfect alignment across different setups reflects how little the tensor-to-scalar ratio correlates with the other parameters.

\begin{figure}[tb]
    % \captionsetup{width=1.2\columnwidth}
    \includegraphics[width=1.15\columnwidth]{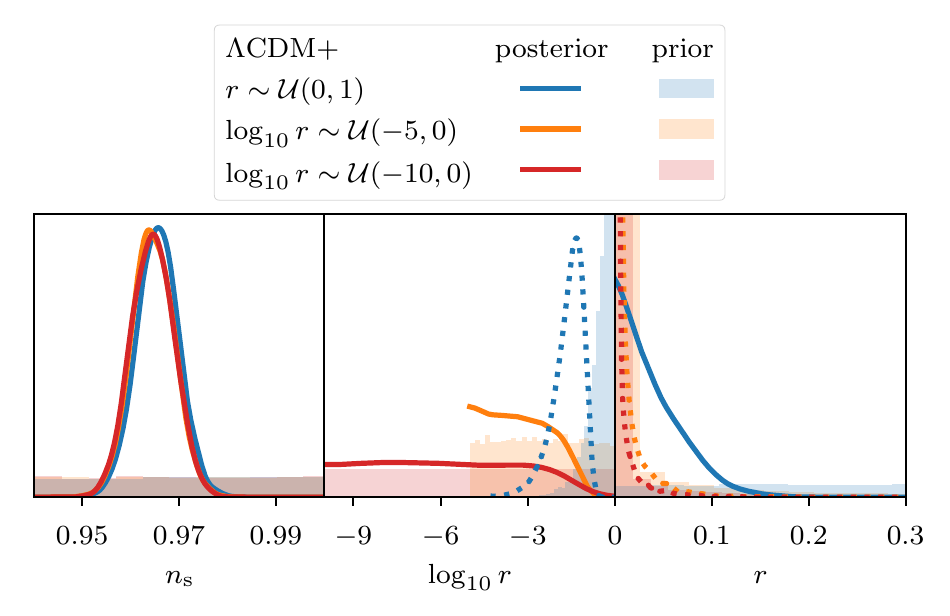}
    \caption{\label{fig:ns_r_1d} Normalised one-dimensional posterior distributions for Planck~2018 TT,TE,EE+lowE data for the spectral index~$n_s$ and the tensor-to-scalar ratio~$r$ of primordial perturbations, contrasting the difference between using a uniform~(blue) or logarithmic~(orange and red) prior on~$r$. The shaded histograms illustrate the prior distributions. Note that the dotted lines show the inferred parameters~$r$ and~$\log_{10}r$ in the respective opposite domain. This is done only to provide a more direct visual comparison. However, these dotted contours are not data-driven parameter constraints. In particular the blue dotted line results purely from a lack of small prior samples when sampling uniformly over~$r$, and does not in fact constitute a lower bound on the tensor-to-scalar ratio.}
\end{figure}

\begin{figure}[tb]
    \flushright
    \includegraphics[]{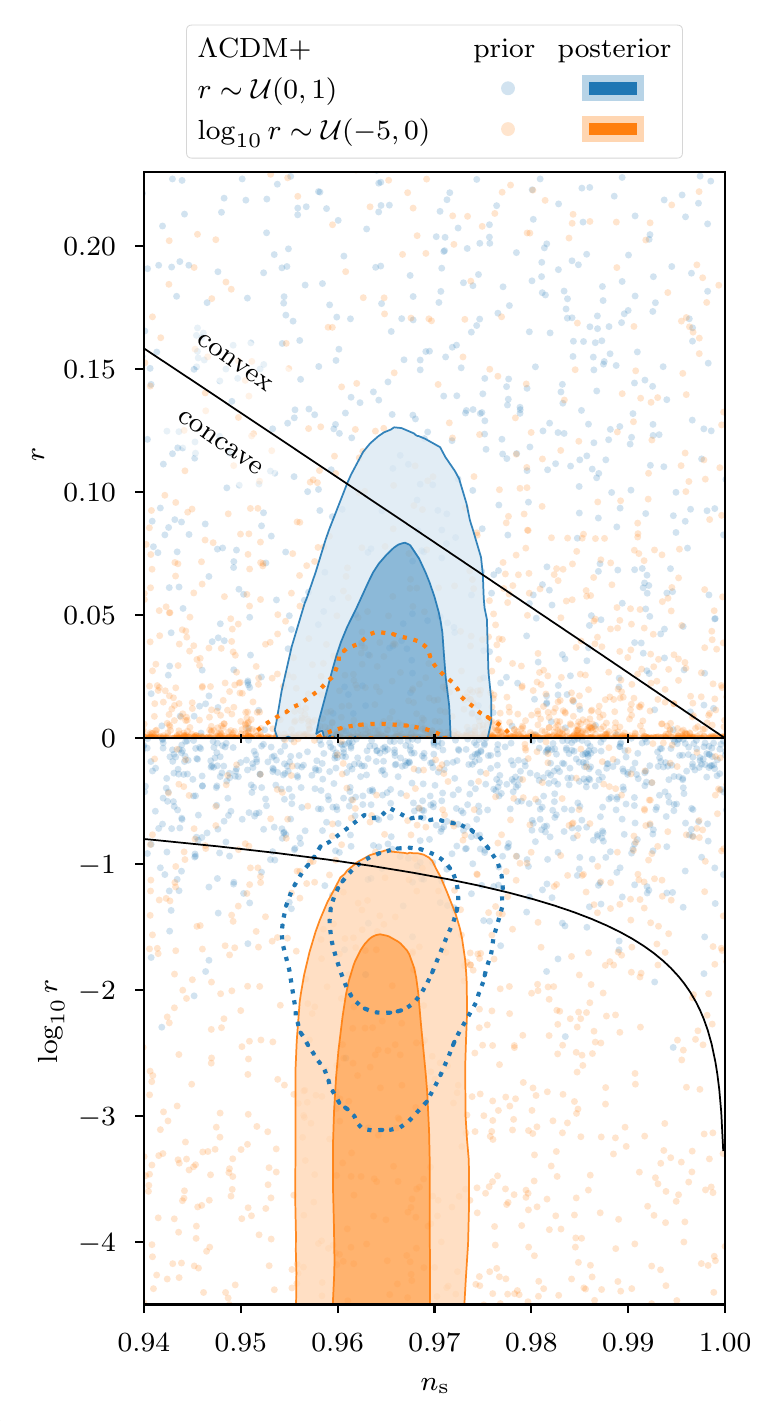}
    \caption{\label{fig:ns_r_2d} Two-dimensional version of \cref{fig:ns_r_1d} showing the \SI{68}{\percent} and \SI{95}{\percent} levels of the posterior contours for Planck~2018 TT,TE,EE+lowE data for spectral index~$n_\mathrm{s}$ and tensor-to-scalar ratio~$r$, where again a uniform~(blue) or logarithmic~(orange) prior on $r$ was used. The scattered dots give a notion of that prior distribution. Note that the dotted lines are not true constraints as explained in \cref{fig:ns_r_1d}. The thin black line divides the $n_\mathrm{s}$-$r$ parameter space into regions of convex and concave inflationary potentials.}
\end{figure}

In \cref{fig:ns_r_1d} we focus on the spectral index~$n_\mathrm{s}$ and the tensor-to-scalar ratio~$r$ (or $\log_{10}r$) in particular by showing their one-dimensional marginalised posterior distributions. \Cref{fig:ns_r_2d} shows the corresponding two-dimensional joint probability contours of the \SI{68}{\percent} and \SI{95}{\percent} levels for $n_\mathrm{s}$ and $r$ (or $\log_{10}r$).
We have included shaded histograms in the 1d plots and scatter points in the 2d plots to give a notion of the prior distributions.

As already expected from \cref{fig:r_param_stability}, the marginalised posterior for the spectral index is near identical, irrespective of the prior on~$r$. The tensor-to-scalar ratio in the right panel of \cref{fig:ns_r_1d} drops off exponentially from $r=0$ to larger values, thereby significantly compressing the prior, which spans up to unity. When sampling logarithmically the posterior levels off towards small scales and shows a step-like behaviour at the upper bound. 

We have included the kernel density estimate from the uniform $r$-samples in the $\log_{10}r$ plot and vice versa (dotted lines). This allows us to compare more directly what sort of numerical values were actually used in those two cases. At a first naive glance one might be concerned that the dotted blue line actually indicates a lower bound, however, looking at the blue shaded histogram in the 1d plot or the blue scatter points in the 2d plot it becomes clear that this is entirely prior driven and reflects that uniform sampling of~$r$ does not reach such low values (see also~\cite{Forbes2020} on a related discussion about the importance of adjusting the density when setting the $x$-scale to `log'). With a target of $r \sim 10^{-3}$ this highlights how the parameter space is sampled rather inefficiently at those low values of interest when applying a uniform prior, which would be an argument for adopting a logarithmic prior in the future. 

One problem to be aware of with the unconstrained posteriors from a logarithmic prior is that upper bounds in form of e.g. \SI{95}{\percent}~limits will change with the lower bound on the logarithmic parameter: the smaller the lower prior bound, the smaller also the upper posterior bound. 
This lack of a stable posterior bound is a result of the definition via percentiles, a notion inspired by a normal distribution. For other types of distributions, such as the step-like posteriors seen in the middle panel of \cref{fig:ns_r_1d}, percentiles of that sort are not the ideal measure for an upper bound. 
For such a step-like posterior a better alternative would be to quantify the position of the step directly, e.g.\ where the posterior drops to some fraction of its plateau value.
In the case that an exponential distribution provides a good fit to the non-logarithmic parameter (see the mock example in the following \cref{sec:mock}), the parameter value where the posterior is~$1/\e$ times its maximum turns out to be a stable choice, which corresponds to the mean of the exponential distribution.
Indeed, using that $1/\e$ measure for the step position we get roughly the same upper bound on the tensor-to-scalar ratio for all prior options:
\begin{align}
\begin{split}
    r &< 0.06 , \\
    \log_{10}r &< -1.2 . 
\end{split}
\end{align}
Note, that these are \emph{not} the habitually quoted \SI{95}{\percent} upper bounds on the tensor-to-scalar ratio. For the uniform sampling run of~$r$, this limit in this case is closer to roughly an \SI{80}{\percent} upper bound. 
Note further that the choice of the $1/\e$ fraction provides a particularly stable bound, because of the connection to the mean of the exponential distribution.

\subsection{Tensor-to-scalar ratio: Evidence and Kullback--Leibler divergence}
\label{sec:r_stats}

\begin{figure*}[tbp]
    \includegraphics[scale=1.1]{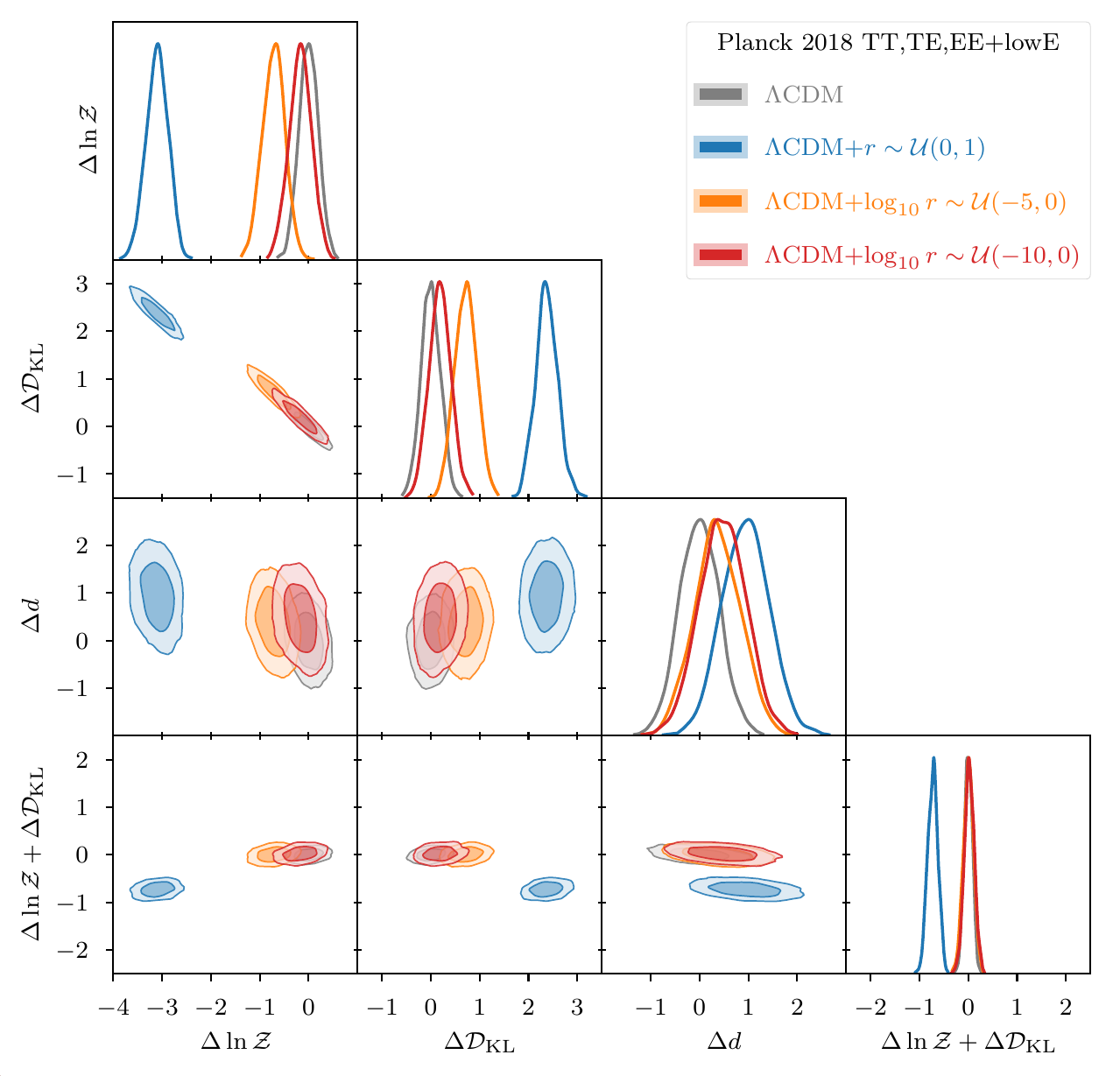}
    % \vspace{-1ex}
    \caption{\label{fig:r_stats} Effect of uniform vs logarithmic priors on Bayesian model comparison for the tensor-to-scalar ratio~$r$: log-evidence~$\Delta\ln\mathcal{Z}$, Kullback--Leibler divergence~$\mathcal{D}_\mathrm{KL}$ (in \si{\nats}), Bayesian model dimensionality~$d$, and posterior average of the log-likelihood~$\langle\ln\mathcal{L}\rangle_\mathcal{P} = \ln\mathcal{Z}+\mathcal{D}_\mathrm{KL}$. 
    The probability distributions represent errors arising from the nested sampling process. In the limit of infinite life points these distributions would become point statistics, in contrast to posterior distributions.
    We normalise with respect to the $\Lambda$CDM model without $r$ (i.e.\ with $r=0$). Note, how switching from uniform to logarithmic sampling of~$r$ (i.e.\ from blue to orange/red) moves the contours along their $\ln\mathcal{Z}$,~$\mathcal{D}_\mathrm{KL}$ degeneracy line, i.e.\ relative entropy is traded in for evidence. Note further by comparison of the orange and red lines, how changing the lower bound of the logarithmic sampling interval (by 5~log-units) barely affects the contours (bar some expected statistical fluctuation due to the sampling error).}
\end{figure*}

\newcolumntype{-}{D{,}{\!\pm\!}{1}}
\begingroup
\squeezetable
\begin{table*}[tb]
\renewcommand{\arraystretch}{1.6}
    \caption{\label{tab:r_stats} Mean and standard deviation of the log-evidence~$\ln\mathcal{Z}$, Kullback--Leibler divergence~$\mathcal{D}_\mathrm{KL}$ and Bayesian model dimensionality~$d$ of the base $\Lambda$CDM cosmology and its $r$~extension from Planck~2018 TT,TE,EE+lowE data~\cite{Planck2018CMB}. The~$\Delta$ indicates normalisation with respect to the base $\Lambda$CDM model.}
\begin{ruledtabular}
    
\begin{tabular}{l------}
Model 
% &\multicolumn{1}{c}{betting odds} 
&\multicolumn{1}{c}{$\ln\mathcal{Z}$} 
&\multicolumn{1}{c}{$\mathcal{D}_\mathrm{KL}$} 
&\multicolumn{1}{c}{$d$} 
&\multicolumn{1}{c}{$\Delta\ln\mathcal{Z}$} 
&\multicolumn{1}{c}{$\Delta\mathcal{D}_\mathrm{KL}$} 
&\multicolumn{1}{c}{$\Delta d$} \\ \hline
$\Lambda$CDM                                                   &-1431.05\pm0.20 &38.57\pm0.20 &17.10\pm0.40 &-0.00\pm0.20 &0.00\pm0.20 &0.00\pm0.40 \\ 
$\Lambda$CDM$+\hphantom{\log_{10}}\,r\sim\mathcal{U}(\;0, 1)$  &-1434.15\pm0.23 &40.94\pm0.23 &18.05\pm0.48 &-3.10\pm0.23 &2.37\pm0.23 &0.95\pm0.48 \\ 
$\Lambda$CDM$+          \log_{10}   r\sim\mathcal{U}( -5, 0)$  &-1431.76\pm0.23 &39.29\pm0.23 &17.47\pm0.47 &-0.71\pm0.23 &0.72\pm0.23 &0.37\pm0.47 \\ 
$\Lambda$CDM$+          \log_{10}   r\sim\mathcal{U}(-10, 0)$  &-1431.22\pm0.23 &38.76\pm0.22 &17.57\pm0.48 &-0.17\pm0.23 &0.19\pm0.22 &0.47\pm0.48 \\ 
%             logZ               D               d             logZ+D       deltalogZ         deltaD          deltad   deltalogZ+deltaD    
% -1431.05 +- 0.20   38.57 +- 0.20   17.10 +- 0.40   -1392.48 +- 0.08    0.00 +- 0.20   0.00 +- 0.20   -0.00 +- 0.40       0.00 +- 0.08    
% -1434.15 +- 0.23   40.94 +- 0.23   18.05 +- 0.48   -1393.21 +- 0.10   -3.10 +- 0.23   2.37 +- 0.23    0.95 +- 0.48      -0.73 +- 0.10    
% -1431.76 +- 0.23   39.29 +- 0.23   17.47 +- 0.47   -1392.48 +- 0.10   -0.71 +- 0.23   0.72 +- 0.23    0.37 +- 0.47       0.00 +- 0.10    
% -1431.22 +- 0.23   38.76 +- 0.22   17.57 +- 0.48   -1392.46 +- 0.10   -0.17 +- 0.23   0.19 +- 0.22    0.47 +- 0.48       0.02 +- 0.10   
\end{tabular}

\end{ruledtabular}
\end{table*}
\endgroup

Nested sampling provides us with distributions for log-evidence~$\ln\mathcal{Z}$, KL-divergence~$\mathcal{D}_\mathrm{KL}$ and Bayesian model dimensionality~$d$ in the same way as for the posterior of free model parameters, which can be calculated straightforwardly using \texttt{anesthetic}'s analysis tools for nested sampling output~\cite{Anesthetic}.
\Cref{fig:r_stats} shows the contours for those quantities in a triangle plot. We have normalised all quantities with respect to the base $\Lambda$CDM model, such that e.g.\ for the log-evidence we have:
\begin{equation}
    \Delta\ln\mathcal{Z} = \ln\mathcal{Z} - \ln\mathcal{Z}_{\Lambda\mathrm{CDM}} .
\end{equation}
\Cref{tab:r_stats} lists the summary statistics for the quantities from \cref{fig:r_stats}.

The marginalised plot for the difference in log-evidence (topmost panel) with $\Delta\ln\mathcal{Z} = -3.10 \pm 0.23$ for the $r$\=/extension of $\Lambda$CDM shows that it is considerably disfavoured when applying a uniform prior. However, switching from a uniform to a logarithmic prior negates the difference in log-evidence completely, such that the $\log r$ extension ends up almost on par with the base $\Lambda$CDM model.

Changing the lower bound for the logarithmic prior, on the other hand, barely affects the evidence value at all. We have performed a run with a lower bound of $\log_{10}r=-5$ and another with $\log_{10}r=-10$, i.e.\ five orders of magnitude difference in~$r$. Despite this large difference in the lower bound the corresponding log-evidence~$\ln\mathcal{Z}$ changes only very little such that the distributions significantly overlap one another. As explained in \cref{sec:bayes}, this is due to $\log r$ being unconstrained below a certain threshold and the Bayesian evidence picking up only on \emph{constrained} parameters. This can seem counter-intuitive, since the Bayesian evidence is generally understood to automatically penalise additional parameters. The key point is that the Occam penalty essentially enters into the Bayesian evidence in the form of the ratio of posterior to prior volume. If both volumes are the same, then they divide out and do not contribute to the Occam penalty.

The last point becomes clearer by also taking into account the KL-divergence and recalling \cref{eq:evidence_fit_occam}, where we identified~$\mathcal{D}_\mathrm{KL}$ as a measure for the Occam penalty.
Looking at the correlation plot between log-evidence and KL-divergence makes it clear that there is a trade-off happening between those two quantities when switching between uniform and logarithmic priors. While the evidence increases for the logarithmic prior, the KL-divergence decreases, as expected from the posterior plots in \cref{fig:ns_r_1d}, which shows how the change from prior to posterior happens only at about $\log_{10}r \gtrsim -2$.
This is further reflected in the Bayesian model dimensionality~$d$, which shows a clear growing trend from about $d=17$ for the base $\Lambda$CDM model via a $\log r$ extension to about $d=18$ for the $r$ extension reflecting the one additional sampling parameter. Note that the total number of sampled parameters consists of 6~base cosmological parameters (+1~for the $r$ extension) and 21 nuisance parameters from the Planck likelihood. 

Because of the trade-off between log-evidence and KL-divergence it is interesting also to look at their sum, which from \cref{eq:evidence_fit_occam} we know turns out to be the posterior \emph{average} of the log-likelihood:
\begin{equation}
    \ln\mathcal{Z} + \mathcal{D}_\mathrm{KL} = \langle\ln\mathcal{L}\rangle_\mathcal{P} .
\end{equation}
This makes for an interesting pairing with the Bayesian model dimensionality, since $d/2$ is the posterior \emph{variance} of the log-likelihood. As such, these two quantities provide an alternative perspective to that of the evidence and KL-divergence.
The posterior average and variance of the log-likelihood are a measure of the fit and complexity respectively.
$\langle \ln\mathcal{L} \rangle_\mathcal{P}$ is shown in the last panel in \cref{fig:r_stats}, where we indeed see that the line for uniform sampling of~$r$ has moved much closer to the other lines, which is to be expected, since~$r$ and~$\log r$ are fundamentally the same parameter and therefore lead to a similar goodness of fit.

This behaviour can also be understood analytically, which we explore in the following section in a one-dimensional mock example, simulating the $r$ vs $\log r$ result.

\subsection{Mock example}
\label{sec:mock}

\begin{figure*}[tb]
    \includegraphics[]{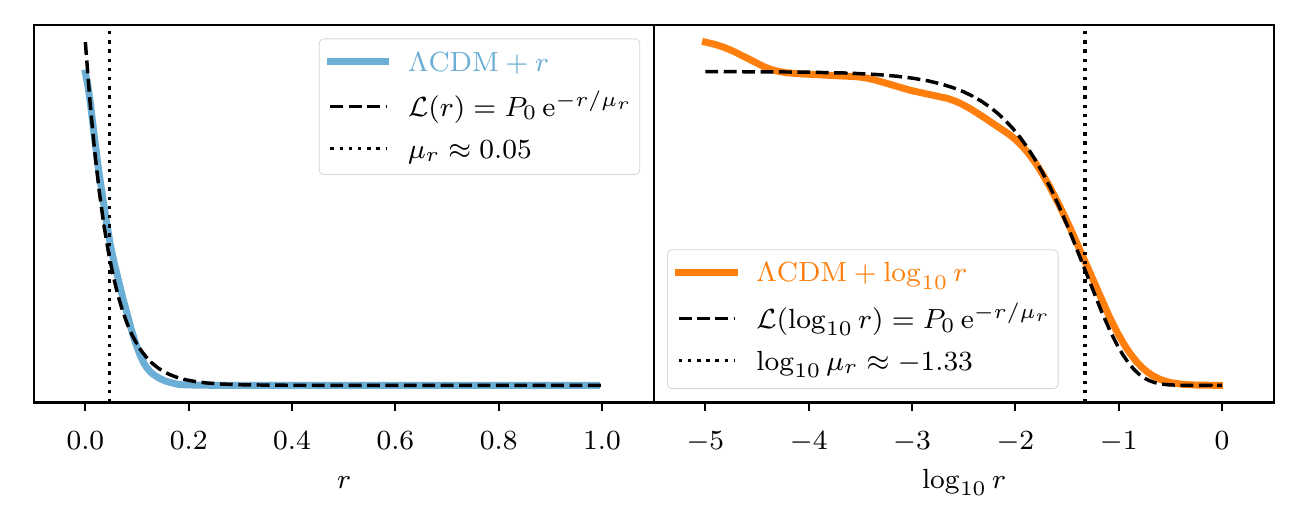}
    % \vspace{-1ex}
    \caption{\label{fig:mock} Exponential likelihood distribution from our mock example in \cref{eq:mockL} compared to Planck~2018 temperature and polarisation data (TT,TE,EE+lowE) on the tensor-to-scalar ratio~$r$ with uniform sampling of~$r$ on the left and logarithmic sampling of~$r$ on the right. Note how the mean~$\mu_r$ fulfills the ordering required by \cref{eq:mock_ordering} and how the lower limit on~$\log_{10}r$ is well into the saturation plateau of posterior/likelihood.}
\end{figure*}

To illustrate further the role of a uniform vs a logarithmic prior on a Bayesian model comparison, we propose the following mock example, which is loosely based on the pedagogical example by~\begin{NoHyper}\citeauthor{Sivia2006ch4}\end{NoHyper}~\cite{Sivia2006ch4} explaining the effect of an additional (although in that case \emph{constrained}) parameter, which we already outlined in \cref{sec:bayes}.

Here, we will not assume a Gaussian likelihood that ultimately fully constrains a parameter, but rather we will assume an exponential distribution as our likelihood on a strictly positive parameter
(which is the maximum-entropy distribution when only a mean is known):
\begin{equation}
\label{eq:mockL}
    \mathcal{L}(a) = P_0 \e^{-a/\mu} ,
\end{equation}
where $P_0=\Pr(D\,|\,a=0)$ is the maximum likelihood value for the data~$D$ at $a=0$ and where~$\mu$ is the mean of the likelihood distribution describing the data. 
Thus, the likelihood is constrained only on one side, providing an upper bound, as shown in the left panel of \cref{fig:mock}.

We will assume a model~$A$, where we sample the parameter~$a$ uniformly in the interval $[a_1, a_2]$. Furthermore, we will assume a model~$B$, where we uniformly sample the parameter~$b=\log_{10}a$ in the interval $[b_1, b_2]$, corresponding to logarithmically sampling the parameter~$a$. 
Since both models are fundamentally governed by the same quantity and will use the same likelihood, any difference in Bayesian inference quantities will be purely prior driven.

We will make the following assumptions on the ordering of the prior limits:
\begin{equation}
\label{eq:mock_ordering}
    0 = a_1 < 10^{b_1} \ll \mu \ll 10^{b_2} = a_2 = 1 .
\end{equation}
This ordering is motivated as follows: For the upper limit we require that the likelihood has essentially dropped to zero. Hence, without loss of generality, we can set the upper limit to one and require $\mu \ll 1$. The lower limit for the positive parameter~$a$ can be explicitly set to zero when sampling uniformly. However, when sampling logarithmically we need to pick some finite lower limit, which we require to be in the region $10^{b_1} \ll \mu$, where the likelihood has essentially saturated with respect to~$b$ (see right panel in \cref{fig:mock}). 
The dependence of Bayesian quantities such as the evidence~$\mathcal{Z}$ or the Kullback--Leibler divergence~$\mathcal{D}_\mathrm{KL}$ on the prior choice on the one hand and on this lower limit~$b_1$ on the other is the goal of this mock example.

The corresponding priors for models~$A$ and~$B$ can thus be written as:
\begin{align}
    \pi_A(a) &=   \frac{1}{a_2-a_1} \,\Theta(a-a_1) \,\Theta(a_2-a) , \\%&
    \pi_B(b) &= \,\frac{1}{b_2-b_1} \;\Theta(b-b_1) \;\Theta(b_2-b) , 
\end{align}
where~$\Theta(x)$ is the Heaviside step function.

\begin{widetext}
We can compute the evidence and Kullback--Leibler divergence for models~$A$ and~$B$ as:
\addtolength{\jot}{9pt}
\begin{align}
    \label{eq:mockZA}
    \mathcal{Z}_A &= \int \, \mathcal{L}(a) \,\, \pi_A(a) \, \dd{a} \,
                  = \frac{P_0 \, \mu}{a_2-a_1} \left( \e^{-a_1/\mu} - \e^{-a_2/\mu} \right) , \\
    \label{eq:mockZB}
    \mathcal{Z}_B &= \int \mathcal{L}(10^b) \, \pi_B(b) \dd{b}
                  = \frac{P_0}{b_2-b_1} \; \frac{1}{\ln(10)} \left[ \mathrm{Ei}\left(-\frac{10^{b_2}}{\mu}\right) - \mathrm{Ei}\left(-\frac{10^{b_1}}{\mu}\right) \right] , \\
    \mathcal{D}_{\mathrm{KL},A} &= \int \; \frac{\mathcal{L}(a) \;\; \pi_A(a)}{\mathcal{Z}_A} \; \ln\left( \frac{\mathcal{L}(a)}{\mathcal{Z}_A} \right) \; \dd{a} \;
                                = \ln\frac{P_0}{\mathcal{Z}_A} - 1 - \frac{P_0}{\mathcal{Z}_A} \; \frac{1}{a_2 - a_1} \left[ a_1 \exp(-\frac{a_1}{\mu}) - a_2 \exp(-\frac{a_2}{\mu}) \right] , \\ 
    \mathcal{D}_{\mathrm{KL},B} &= \int \frac{\mathcal{L}(10^b) \, \pi_B(b)}{\mathcal{Z}_B} \ln\left( \frac{\mathcal{L}(10^b)}{\mathcal{Z}_B} \right) \dd{b} 
                                = \ln\frac{P_0}{\mathcal{Z}_B} - \frac{P_0}{\mathcal{Z}_B} \; \frac{1}{b_2 - b_1} \; \frac{1}{\ln(10)} \left[ \exp(-\frac{10^{b_1}}{\mu}) - \exp(-\frac{10^{b_2}}{\mu}) \right] ,
\end{align}
\setlength{\jot}{3pt}
where $\mathrm{Ei}$ refers to the exponential integral.
With the ordering from \cref{eq:mock_ordering} we can then approximate these to give:
% \begingroup
% \addtolength{\jot}{6pt}
\begin{align}
\label{eq:lnZA}
    &&&&\Delta\ln\mathcal{Z}_A &\approx \ln\mu       \vphantom{\frac11}            &&\sim -3 , &&&&&&\\
\label{eq:lnZB}
    &&&&\Delta\ln\mathcal{Z}_B &\approx \ln\left(1-\frac{\log_{10}\mu}{b_1}\right) &&\sim  0 , &&&&&&\\
\label{eq:DA}
    &&&&\Delta\mathcal{D}_{\mathrm{KL},A} &\approx-\Delta\ln\mathcal{Z}_A-1 \vphantom{\frac11} &&\sim 2 , &&&&&&\\
\label{eq:DB}
    &&&&\Delta\mathcal{D}_{\mathrm{KL},B} &\approx-\Delta\ln\mathcal{Z}_B \vphantom{\frac11} &&\sim 0 , &&&&&& 
\end{align}
% \setlength{\jot}{3pt}
% \endgroup
\end{widetext}
where we normalise with respect to a base model~$O$ with $a=10^b=0$ fixed, such that $\mathcal{Z}_O=P_0$ and $\mathcal{D}_{\mathrm{KL},O}=0$. The numerical values assume $\mu \sim 0.06$, which is roughly the posterior mean of the tensor-to-scalar ratio under uniform sampling in the preceding section. Hence, we can compare these zeroth-order numerical approximations to the results in \cref{fig:r_stats,tab:r_stats}, which indeed match. 

\begin{figure}[tb]
    \includegraphics[width=\columnwidth]{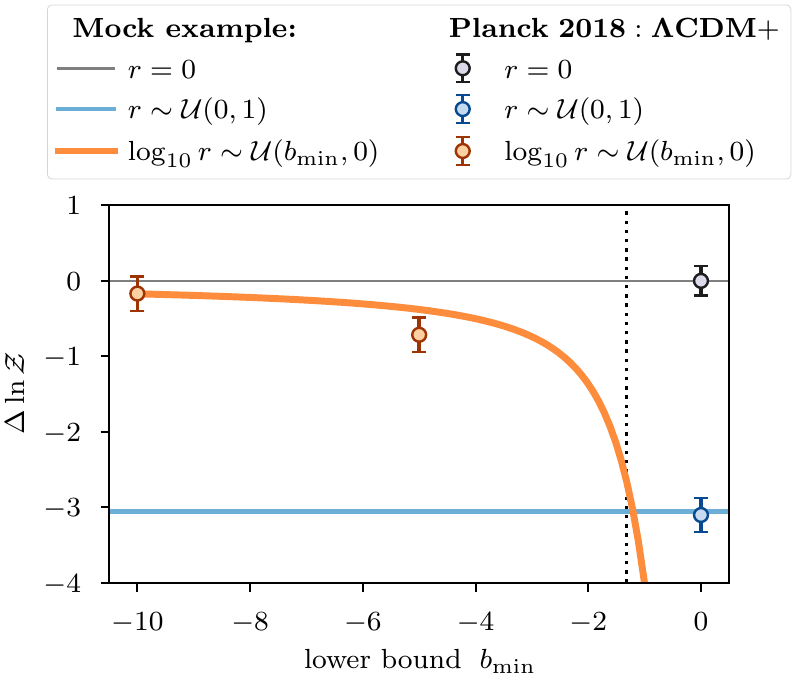}
    \caption{\label{fig:mock_lower_bound} Dependence of the log-evidence on the lower prior bound~$b_\mathrm{min}$: Comparison of the results in \cref{eq:mockZA,eq:mockZB} for the one-dimensional mock example (solid lines) to the nested sampling results from \cref{tab:r_stats} (dots with error bars). The vertical dotted line corresponds to the mean used in the mock likelihood distribution (cf.\ \cref{fig:mock}).}
\end{figure}

\Cref{fig:mock_lower_bound} makes this comparison more thoroughly, comparing the results from our one-dimensional mock example in \cref{eq:mockZA,eq:mockZB} with the nested sampling results from \cref{tab:r_stats} for a variable lower bound~$b_\mathrm{min}$ of the logarithmic prior. 
The mean $\ln\mathcal{Z}$ of the base model with $r=0$ for both the mock example and for the base $\Lambda$CDM nested sampling run are zero by definition of our normalisation. They serve only as calibration for the models with uniform~(blue) and logarithmic~(orange) priors. All three nested sampling runs agree well with the prediction from the mock example within their margins of errors.

\Cref{fig:mock_lower_bound} illustrates how the evidence levels off with regards to the choice of the lower bound of the logarithmic prior (orange line) also reflected in the near equal evidences of the nested sampling runs with lower prior bounds of~$-5$ and~$-10$ respectively.
Note that the good agreement between mock example and data in \cref{fig:mock_lower_bound} is due to the fact that the tensor-to-scalar ratio is almost completely uncorrelated with the other cosmological parameters, with the biggest (yet still small) correlation coming from the spectral index~$n_\mathrm{s}$ (cf.\ \cref{fig:ns_r_2d}).

\section{Neutrino masses}
\label{sec:mnu}

\begin{figure*}[tbp]
    \includegraphics[]{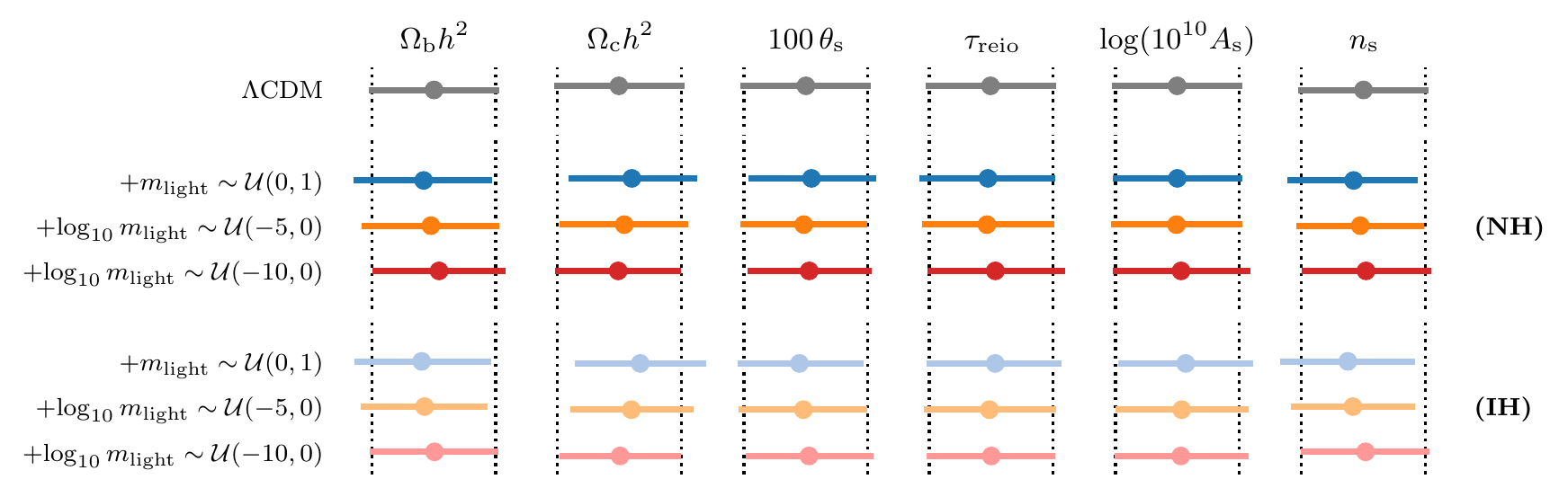}
    \caption{\label{fig:nu_param_stability} Stability of the cosmological parameters for the 3-neutrino extension of the base $\Lambda$CDM cosmology for different priors on $m_\mathrm{light}$: uniform in blues, logarithmic with lower bound of~$-5$ in oranges and logarithmic with lower bound of~$-10$ in reds. The darker set of colours corresponds to the normal neutrino hierarchy~(NH) and the lighter set to the inverted hierarchy~(IH). For each parameter we show the mean and the extent from quantile $0.16$ to $0.84$, i.e.\ the inner \SI{68}{\percent} limits.}
\end{figure*}

In Planck's baseline cosmology, the neutrinos are assumed to be comprised of two massless neutrinos and one massive neutrino with mass~$m_\nu=\SI{0.06}{\eV}$ with the effective number of neutrino species set slightly larger than~3 to $N_\mathrm{eff}=3.046$~\cite{Planck2018Parameters,Mangano2005,deSalas2016}. 

Upcoming CMB experiments such as the Simons Observatory, LiteBIRD or CMB\=/S4 and large scale structure~(LSS) experiments such as Euclid will allow us to fully constrain the sum of neutrino masses~$\sum m_\nu$. However, even under the most optimistic assumptions, it will not be possible to disentangle the individual contributions of the three neutrino flavours with cosmological data alone~\cite{Archidiacono2020}. 
To achieve that, we need additional data from solar, atmospheric, reactor and accelerator experiments as summarised in NuFIT~5.0~(2020)~\cite{Esteban2020,NuFIT} that provide us with the mass square splittings:
\begin{align}
\label{eq:delta_m2}
    \delta m^2 &= 7.42_{-0.20}^{+0.21} \times \SI{e-5}{\eV\squared} \quad\; \text{(NH \& IH)} , \\
\label{eq:Delta_m2}
    \Delta m^2 &= 
\begin{cases}
    2.517_{-0.028}^{+0.026} \times \SI{e-3}{\eV\squared} &\text{(NH)} , \\
    2.498_{-0.028}^{+0.028} \times \SI{e-3}{\eV\squared} &\text{(IH)} , 
\end{cases}
\end{align}
where $\delta m^2$ is the smaller squared mass splitting between the light and the medium neutrino mass for the normal neutrino hierarchy~(NH) and between the medium and the heavy neutrino mass for the inverted neutrino hierarchy~(IH), and $\Delta m^2$ is the larger squared mass splitting between the light and the heavy neutrino mass in both cases.

With the knowledge of the two squared mass splittings, the remaining uncertainty lies mostly with the scale of the lightest neutrino. In the following Bayesian analysis we therefore apply Gaussian priors according to \cref{eq:delta_m2,eq:Delta_m2} and vary over the lightest neutrino mass.

\subsection{Neutrino masses: Posteriors}
\label{sec:nu_posterior}

\begin{figure*}[tbp]
    \vspace{1cm}
    \subfloat{\includegraphics[scale=1.1]{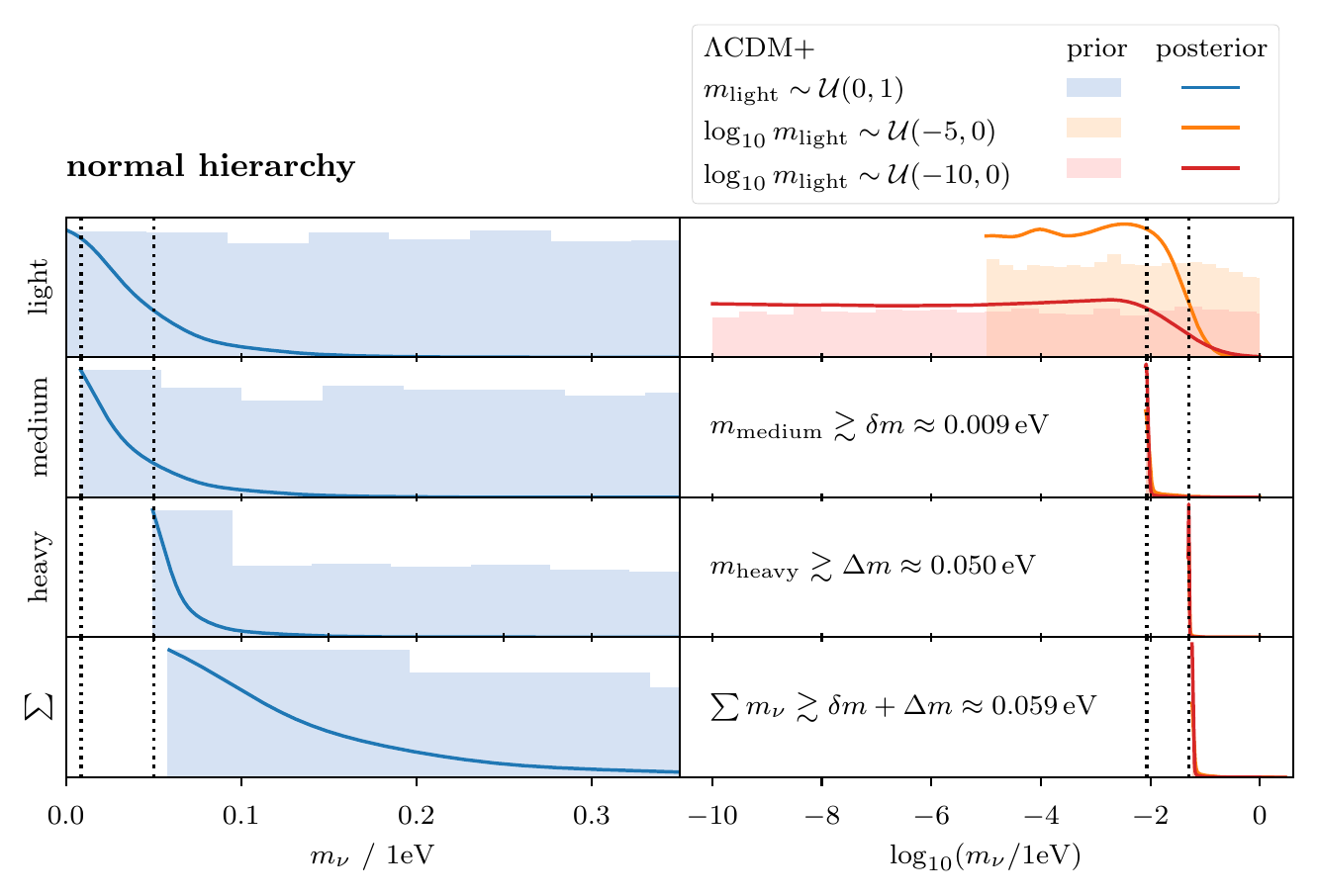}} \\
    \subfloat{\includegraphics[scale=1.1]{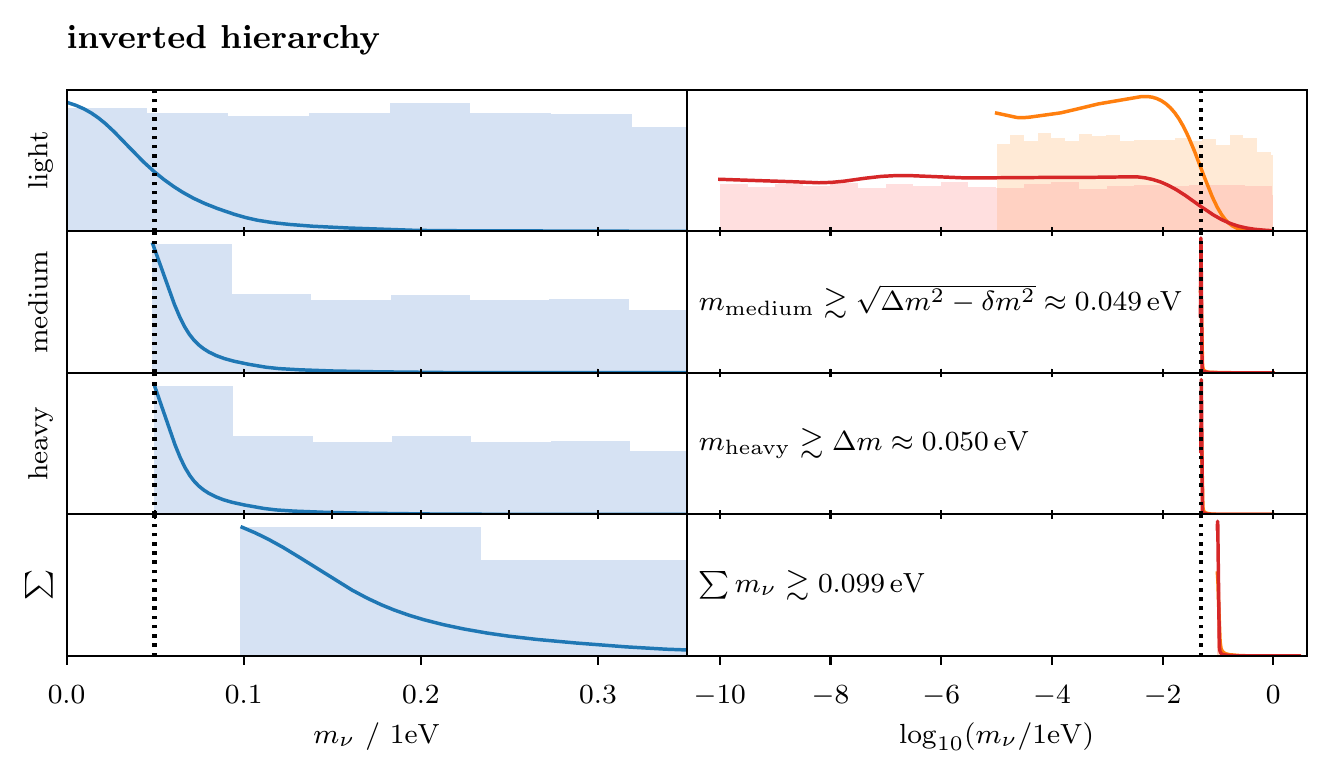}}
    \caption{\label{fig:nu_posterior} One-dimensional posterior distributions of neutrino masses with normal hierarchy~(NH) in the top panel and with inverted hierarchy~(IH) in the bottom panel for TT,TE,EE+lowE data from Planck~2018 and neutrino oscillation data on the mass squared splittings from NuFIT~5.0~(2020). The vertical black dotted lines give the rough lower limit on medium and heavy mass that is set by the mass squared splittings~$\delta m^2$ and~$\Delta m^2$. For the inverted hierarchy these dotted lines appear almost on top of each other. The rows show the posteriors for the light, medium, and heavy neutrino mass and sum of all neutrino masses, respectively. The columns contrast the difference between using a uniform~(blue, left) or logarithmic~(orange and red, right) prior on the light neutrino mass~$m_\mathrm{light}$. The shaded histograms give a notion of that prior distribution.}
\end{figure*}

We have taken nested sampling runs for an extension of the base $\Lambda$CDM cosmology with three individual neutrino masses, where we have used both a uniform prior $m_\mathrm{light}\sim\mathcal{U}(0,1)$ and logarithmic priors with different lower bounds, $\log_{10}m_\mathrm{light}\sim\mathcal{U}(-5,0)$ and $\log_{10}m_\mathrm{light}\sim\mathcal{U}(-10,0)$, on the lightest neutrino mass. The other two neutrino masses are then derived from $m_\mathrm{light}$ together with~$\delta m^2$ and~$\Delta m^2$ from \cref{eq:delta_m2,eq:Delta_m2}:
\begin{align}
    m_\mathrm{medium}^2 &= 
    \begin{cases}
        m_\mathrm{light}^2 + \delta m^2              & \mathrm{(NH)}, \\
        m_\mathrm{light}^2 + \Delta m^2 - \delta m^2 & \mathrm{(IH)},    
    \end{cases} \\[6pt]
    m_\mathrm{heavy}^2 &= m_\mathrm{light}^2 + \Delta m^2 .
\end{align}

\Cref{fig:nu_param_stability} gives an overview of the stability of the cosmological base parameters across the different priors for~$m_\mathrm{light}$ and compares them to the $\Lambda$CDM base model by showing their mean and \SI{68}{\percent} ranges.
Compared to \cref{fig:r_param_stability} for the tensor-to-scalar ratio there are some small parameter shifts visible in relation to the base $\Lambda$CDM model, but all shifts stay well within the \SI{68}{\percent} bounds. 

\Cref{fig:nu_posterior} shows the one-dimensional marginalised posterior distributions for the three individual neutrino masses~$m_\mathrm{light}$, $m_\mathrm{medium}$, and~$m_\mathrm{heavy}$, as well as the sum of all three~$\sum m_\nu$ for both the normal and the inverted hierarchy. We have included shaded histograms to give a notion of the prior distributions. The vertical black dotted lines indicate roughly the lower bound for the medium and heavy neutrino mass as determined from the mass squared splittings under the assumption where the light neutrino mass is zero.

When looking at the lightest neutrino mass in the first row, the picture is very similar to that for the tensor-to-scalar ratio before, and most of what we have said in \cref{sec:r_posterior} applies here, too. One has an almost exponential drop-off from zero when sampling uniformly over the mass (left column), significantly compressing the prior, which turns into a more step-like behaviour with respect to the logarithm of the mass when sampling the mass logarithmically (right column). 

Note that the medium and heavy mass from rows 2 and 3 as well as the sum of all masses in the bottom row are \emph{derived} quantities and therefore do not show the same prior behaviour visible for the light neutrino mass. This is not so apparent for the derived masses, when sampling uniformly over the light neutrino mass, although one can see a slight step in the histogram of the prior for the heavy mass in the NH case and for both medium and heavy mass in the IH case. However, when sampling logarithmically over the light neutrino mass, then the picture is much clearer. The probability density for medium and heavy neutrino mass bulks up around their rough lower minimum set by the smaller and larger mass square splitting respectively.

There are two perspectives that one can adopt here. On one hand, one could criticise the choice of a logarithmic prior for being ultimately too prior (or theory) driven and not reflective of the data. On the other hand, one could say that this is the natural result of our state of knowledge of the mass square splittings and our true ignorance about the scale of the lightest neutrino mass. 

We wonder whether this very last statement could be contested, e.g.\ could we say that we would expect the lightest neutrino mass to be of a magnitude similar to that of the medium neutrino mass in the NH? However, this is not the case, when checking for precedence by looking at the other set of leptons, the electron, muon and tauon, where we have roughly around 2 orders of magnitude between their masses~\cite{ParticleBooklet2020}.

Comparing the two hierarchies with one another, we can see that the major difference lies in the medium neutrino mass (and therefore also the sum of all neutrino masses), which is restricted to larger masses in the inverted hierarchy compared to the normal hierarchy, as expected from the mass square splitting (black dotted lines).

\begin{figure*}[tbp!]
    \includegraphics[scale=1.05]{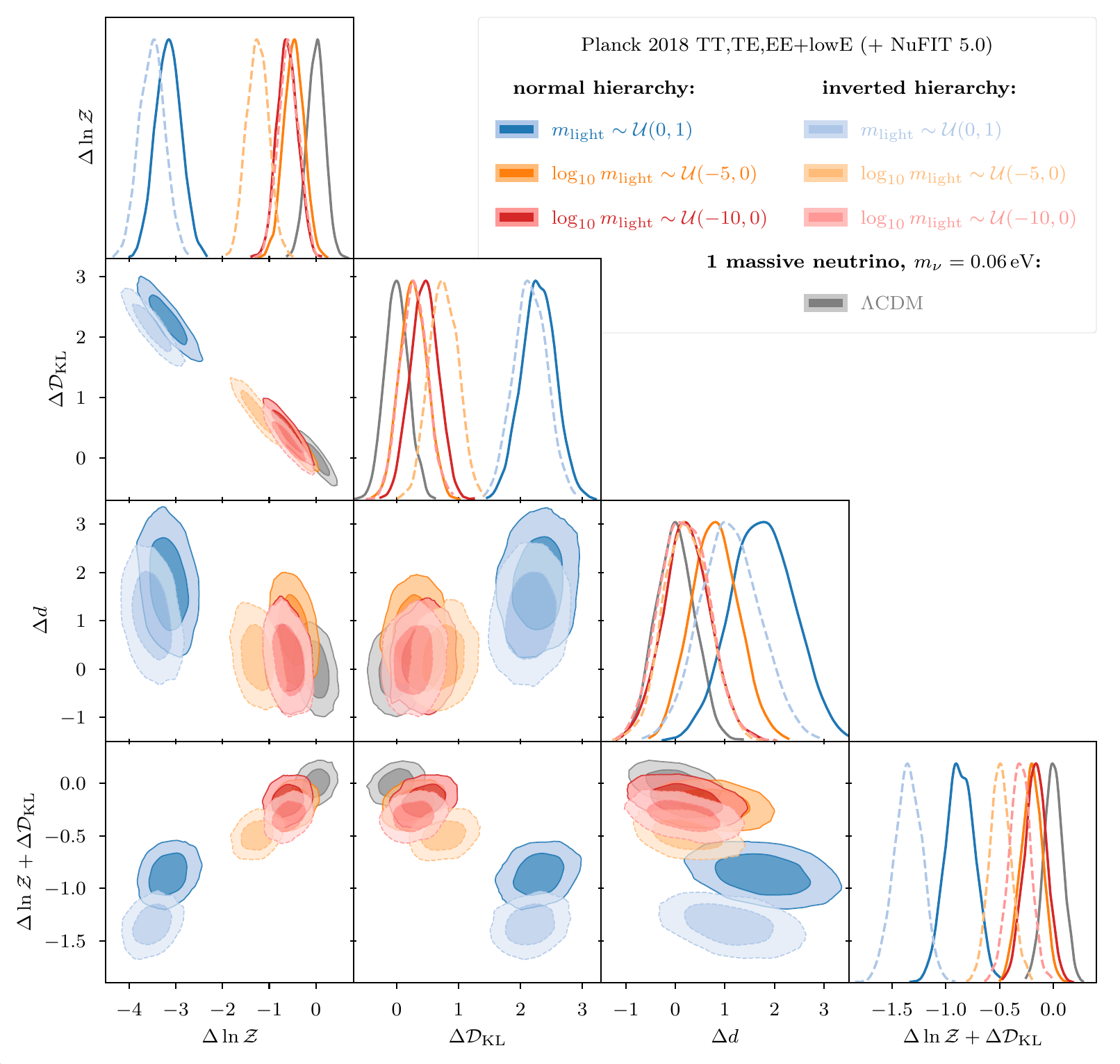}
    \caption{\label{fig:nu_stats} Effect of uniform vs logarithmic priors on the light neutrino mass~$m_\mathrm{light}$ for Bayesian model comparison: log-evidence~$\Delta\ln\mathcal{Z}$, Kullback--Leibler divergence~$\mathcal{D}_\mathrm{KL}$, Bayesian model dimensionality~$d$, and posterior average of the log-likelihood~$\langle\ln\mathcal{L}\rangle_\mathcal{P} = \ln\mathcal{Z}+\mathcal{D}_\mathrm{KL}$. 
    The probability distributions represent errors arising from the nested sampling process. In the limit of infinite life points these distributions would become point statistics, in contrast to posterior distributions.
    We normalise with respect to the $\Lambda$CDM model with two massless and only one massive neutrino with~$m_\nu=\SI{0.06}{\eV}$. Note, how switching from uniform to logarithmic sampling of~$m_\mathrm{light}$ moves the contours along their $\ln\mathcal{Z}$,~$\mathcal{D}_\mathrm{KL}$ degeneracy line, i.e.\ relative entropy is traded in for evidence. Note further by comparison of the orange and red lines, how changing the lower bound of the logarithmic sampling interval (by 5~log-units!) barely affects the contours (bar some expected statistical fluctuation due to the sampling error).}
\end{figure*}

\begingroup
\squeezetable
\begin{table*}[tb]
\renewcommand{\arraystretch}{1.6}
    \caption{\label{tab:nu_stats} Mean and standard deviation of the log-evidence~$\ln\mathcal{Z}$, Kullback--Leibler divergence~$\mathcal{D}_\mathrm{KL}$ and Bayesian model dimensionality~$d$ of the base $\Lambda$CDM cosmology and its 3-neutrino extension from Planck~2018 TT,TE,EE+lowE data~\cite{Planck2018CMB}. The second block of rows shows the results from the normal neutrino hierarchy and the third block for the inverted hierarchy. 
    The~$\Delta$ indicate normalisation with respect to the base $\Lambda$CDM model.}
\begin{ruledtabular}
    
\begin{tabular}{ll------}
&Model 
&\multicolumn{1}{c}{$\ln\mathcal{Z}$} 
&\multicolumn{1}{c}{$\mathcal{D}_\mathrm{KL}$} 
&\multicolumn{1}{c}{$d$} 
&\multicolumn{1}{c}{$\Delta\ln\mathcal{Z}$} 
&\multicolumn{1}{c}{$\Delta\mathcal{D}_\mathrm{KL}$} 
&\multicolumn{1}{c}{$\Delta d$} \\ \hline
                                                    &$\Lambda$CDM                                                                  &-1431.04\pm0.19 &38.56\pm0.19 &17.08\pm0.39 &-0.00\pm0.19 &0.00\pm0.19 &0.00\pm0.39  \\ \hline
\multirow{3}{*}{\rotatebox[origin=c]{90}{normal}}   &$\Lambda$CDM$+\hphantom{\log_{10}}\,m_\mathrm{light}\sim\mathcal{U}(\;0, 1)$  &-1434.20\pm0.27 &40.85\pm0.27 &18.85\pm0.63 &-3.16\pm0.27 &2.29\pm0.27 &1.77\pm0.63  \\
                                                    &$\Lambda$CDM$+          \log_{10}   m_\mathrm{light}\sim\mathcal{U}( -5, 0)$  &-1431.51\pm0.22 &38.83\pm0.22 &17.88\pm0.48 &-0.47\pm0.22 &0.27\pm0.22 &0.80\pm0.48  \\
                                                    &$\Lambda$CDM$+          \log_{10}   m_\mathrm{light}\sim\mathcal{U}(-10, 0)$  &-1431.65\pm0.22 &39.01\pm0.22 &17.31\pm0.48 &-0.61\pm0.22 &0.45\pm0.22 &0.23\pm0.48  \\ \hline
\multirow{3}{*}{\rotatebox[origin=c]{90}{inverted}} &$\Lambda$CDM$+\hphantom{\log_{10}}\,m_\mathrm{light}\sim\mathcal{U}(\;0, 1)$  &-1434.55\pm0.27 &40.72\pm0.27 &18.19\pm0.60 &-3.51\pm0.27 &2.16\pm0.27 &1.11\pm0.60  \\
                                                    &$\Lambda$CDM$+          \log_{10}   m_\mathrm{light}\sim\mathcal{U}( -5, 0)$  &-1432.29\pm0.23 &39.32\pm0.23 &17.31\pm0.46 &-1.25\pm0.23 &0.76\pm0.23 &0.23\pm0.46  \\
                                                    &$\Lambda$CDM$+          \log_{10}   m_\mathrm{light}\sim\mathcal{U}(-10, 0)$  &-1431.63\pm0.22 &38.84\pm0.22 &17.27\pm0.49 &-0.59\pm0.22 &0.28\pm0.22 &0.19\pm0.49  \\
%             logZ                  D                  d             logZ+D          deltalogZ             deltaD             deltad   deltalogZ+deltaD   
% 
% -1431.04 +- 0.19      38.56 +- 0.19      17.08 +- 0.39   -1392.48 +- 0.09       0.00 +- 0.19      -0.00 +- 0.19      -0.00 +- 0.39       0.00 +- 0.09   
% 
% -1434.20 +- 0.27      40.85 +- 0.27      18.85 +- 0.63   -1393.34 +- 0.13      -3.16 +- 0.27       2.29 +- 0.27       1.77 +- 0.63      -0.87 +- 0.13   
% -1431.51 +- 0.22      38.83 +- 0.22      17.88 +- 0.48   -1392.68 +- 0.10      -0.47 +- 0.22       0.27 +- 0.22       0.80 +- 0.48      -0.20 +- 0.10   
% -1431.65 +- 0.22      39.01 +- 0.22      17.31 +- 0.48   -1392.64 +- 0.10      -0.61 +- 0.22       0.45 +- 0.22       0.23 +- 0.48      -0.17 +- 0.10   
% 
% -1434.55 +- 0.27      40.72 +- 0.27      18.19 +- 0.60   -1393.83 +- 0.13      -3.51 +- 0.27       2.16 +- 0.27       1.11 +- 0.60      -1.35 +- 0.13   
% -1432.29 +- 0.23      39.32 +- 0.23      17.31 +- 0.46   -1392.97 +- 0.10      -1.25 +- 0.23       0.76 +- 0.23       0.23 +- 0.46      -0.49 +- 0.10   
% -1431.63 +- 0.22      38.84 +- 0.22      17.27 +- 0.49   -1392.79 +- 0.10      -0.59 +- 0.22       0.28 +- 0.22       0.19 +- 0.49      -0.31 +- 0.10   
\end{tabular}

\end{ruledtabular}
\end{table*}
\endgroup

\subsection{Neutrino masses: Evidence and Kullback--Leibler divergence}
\label{sec:nu_stats}

In \cref{fig:nu_stats} we show the results from our nested sampling runs for the log-evidence~$\ln\mathcal{Z}$, KL-divergence~$\mathcal{D}_\mathrm{KL}$, Bayesian model dimensionality~$d$ and posterior average of the log-likelihood~$\langle \ln\mathcal{L} \rangle$. We again normalise with respect to the base $\Lambda$CDM model. \Cref{tab:nu_stats} lists the summary statistics for these quantities.
As already the case for the posterior, the picture here is again similar to the one for the tensor-to-scalar ratio in \cref{sec:r_stats}.

Looking at the distributions for the log-evidence (topmost diagonal panel) shows that the addition of the neutrino parameters with uniform sampling over the light neutrino mass (either hierarchy) is disfavoured with over 3~log-units compared to the base $\Lambda$CDM model with a single massive neutrino of fixed mass (and 2 massless).
Since the mass square splittings enter on the prior level in our analysis and remain essentially unconstrained by the cosmological data, any change to the evidence is almost entirely driven by the light neutrino mass parameter. Hence, it is not surprising that upon switching to a logarithmic prior on~$m_\mathrm{light}$ the log-evidence increases again while the KL-divergence drops close to the level of the $\Lambda$CDM model. We need to keep in mind that since this is an extension to the $\Lambda$CDM model, it has in principle a better chance of fitting the data, such that any difference in the Bayesian evidence can be attributed to an Occam penalty, which the shift between uniform and logarithmic sampling confirms.

As expected from our investigations for the tensor-to-scalar ratio and especially with regards to our mock example from \cref{sec:mock}, changing the lower bound for the logarithmic prior does not affect the Bayesian evidence. We have again performed runs with two different lower bounds of $\log m_\mathrm{light}=-5$ and $\log m_\mathrm{light}=-10$, i.e.\ five orders of magnitude apart. With both of these bounds well into the area of the posterior (see top right panel of \cref{fig:nu_posterior}) where it has levelled off, we do not expect much change to the evidence value. This is clearly confirmed in \cref{tab:nu_stats,fig:nu_stats} for the normal hierarchy. For the inverted hierarchy it is not as clear but still reasonable in light of the uncertainties. 

Looking at the normalised posterior average of the log-likelihood $\langle\ln\mathcal{L}\rangle_\mathcal{P}=\ln\mathcal{Z}+\mathcal{D}_\mathrm{KL}$ we again roughly confirm 
\begin{align}
    \Delta\ln\mathcal{Z}_\mathrm{uni} + \Delta\mathcal{D}_\mathrm{KL,uni} &\approx -1 , \\
    \Delta\ln\mathcal{Z}_\mathrm{log} + \Delta\mathcal{D}_\mathrm{KL,log} &\approx 0 ,
\end{align}
matching our mock results from \cref{eq:lnZA,eq:lnZB,eq:DA,eq:DB}, independent from the mock parameter~$\mu$.

\subsection{Neutrino hierarchy}

A Bayesian model comparison of the normal vs the inverted neutrino hierarchy is beyond the scope of this paper and has been done before with more stringent data~\cite{Hannestad2016,Heavens2018,Choudhury2020}. However, with posteriors and evidences at hand, we shall briefly discuss the situation here.

There have been claims to a strong preference of the normal over the inverted neutrino hierarchy~\cite{Simpson2017}, however, such strong evidence can typically be traced back to prior volume effects~\cite{Schwetz2017}, i.e.\ the effect of a reduced sampling space for the inverted hierarchy. In other words, we need to watch out and properly distinguish to what extent any Bayesian preference is assigned already on the prior level and to what extent is that preference indeed driven by the data. 

In our analysis both hierarchies start out on an equal footing. With the same prior on the light neutrino mass and equivalent Gaussian priors on the mass squared splittings from neutrino oscillation experiments, the prior volume for both hierarchies is essentially the same. Note that although the means for the larger mass squared splitting~$\Delta m^2$ are slightly different in the two hierarchies, its standard deviations are essentially the same.

There is a slight tendency for all prior options of a better fit of the normal compared to the inverted neutrino hierarchy. However, with an evidence difference of less than one log-unit (odds of maximally $2:1$) any preference for the normal hierarchy is meagre at best, especially when also accounting for the sampling error (see \cref{fig:nu_stats}).
It should be noted, though, that we have used only CMB temperature and polarisation data here. Adding data from CMB lensing or baryon acoustic oscillations would have further shrunk the constraints on the sum of neutrino masses and thereby possibly strengthened the case for the normal hierarchy.

\section{Discussion}
\label{sec:discussion}

We demonstrate how switching between a uniform and a logarithmic prior on some single-bounded model parameter results in a trade-off between Bayesian evidence and Kullback--Leibler divergence (or relative entropy). The common scenario is that of insufficient data sensitivity, leading to a one-sided bound on a parameter. For a location parameter this typically causes an exponential drop-off, which translates to a step-like behaviour when turned into the corresponding scale parameter. We show that the ambiguity of the lower bound of the scale parameter does not affect a Bayesian model comparison, provided the lower bound is chosen sufficiently far into the likelihood plateau.

We demonstrate this behaviour for two cases of parameter extensions to the $\Lambda$CDM model of cosmology, namely for the tensor-to-scalar ratio of primordial perturbations and for the case of three non-degenerate neutrino masses. In both cases we confirm that switching from a uniform prior to a logarithmic prior will get rid of (most of) the Occam penalty associated with that parameter, since unconstrained parameters do not affect the Bayesian evidence. Thus the Bayesian evidence is roughly on par with the un-extended (base) model, with the only difference in the form of an uninformative parameter. Furthermore and for the same reason, the exact choice of the lower bound for the logarithmic prior does not change the Bayesian evidence. When the likelihood levels off, e.g.\ due to insufficient sensitivity in the data, then so does the Bayesian evidence.

\section*{}

\begin{acknowledgments}
LTH would like to thank the Isaac Newton Trust, the STFC, and the Cavendish Laboratory for their support. WJH was supported by a Gonville \& Caius Research Fellowship.

This work was performed using the resources provided by the Cambridge Service for Data Driven Discovery~(CSD3) operated by the University of Cambridge Research Computing Service (www.csd3.cam.ac.uk), provided by Dell EMC and Intel using Tier-2 funding from the Engineering and Physical Sciences Research Council (capital grant EP/P020259/1), and DiRAC (www.dirac.ac.uk) funding from the Science and Technology Facilities Council~(STFC) (capital grants ST/P002307/1 and ST/R002452/1 and operations grant ST/R00689X/1). 
DiRAC is part of the National e-Infrastructure.

This work was also performed using the DiRAC Data Intensive service at Leicester~(DiaL), operated by the University of Leicester IT Services, which forms part of the STFC DiRAC HPC Facility (www.dirac.ac.uk). 
The equipment was funded by BEIS capital funding via STFC (capital grants ST/K000373/1 and ST/R002363/1 and operations grant ST/R001014/1). 
DiRAC is part of the National e-Infrastructure.
\end{acknowledgments}

\bibliography{references_unilog}% Produces the bibliography via BibTeX.

\end{document}